\documentclass[onecolumn,12pt,eqsecnum]{revtex4}
\usepackage{amsmath,amssymb,bm,aas_macros}
\usepackage{natbib}
\bibpunct{[}{]}{;}{a}{}{,}

\usepackage{hyperref}
\begin{document}

\title{An Extension of the Faddeev-Jackiw Technique to Fields in Curved Spacetimes}
\author{C. Prescod-Weinstein}
\email{chanda@mit.edu}
\author{Edmund Bertschinger}
\email{edbert@mit.edu}

\affiliation{Department of Physics \& \\
 Kavli Institute for Astrophysics and Space Research\\
Massachusetts Institute of Technology\\
77 Massachusetts Avenue, MA, 02139 \\
U.S.A.}

\date{\today}

\begin{abstract}
The Legendre transformation on singular Lagrangians, e.g. Lagrangians representing gauge theories, fails due to the presence of constraints. The Faddeev-Jackiw technique, which offers an alternative to that of Dirac, is a symplectic approach to calculating a Hamiltonian paired with a well-defined initial value problem when working with a singular Lagrangian. This phase space coordinate reduction was generalized by Barcelos-Neto and Wotzasek to simplify its application. We present an extension of the Faddeev-Jackiw technique for constraint reduction in gauge field theories and non-gauge field theories that are coupled to a curved spacetime that is described by General Relativity. A major difference from previous formulations is that we do not explicitly construct the symplectic matrix, as that is not necessary.  We find that the technique is a useful tool that avoids some of the subtle complications of the Dirac approach to constraints. We apply this formulation to the Ginzburg-Landau action and provide a calculation of its Hamiltonian and Poisson brackets in a curved spacetime.
\end{abstract}

\maketitle

\section{Introduction}\label{intro}
The usual approach toward finding the Hamiltonian or symplectic structure of a dynamical system is to begin with the Lagrangian, perform a Legendre transformation, and assume Poisson brackets. In many cases, this is enough. However, when the Lagrangian is singular, this approach will fail. A singular Lagrangian describes a constrained system -- there are more degrees of freedom in the resulting equations of motion that are derived from the Euler-Lagrange equations than the number of physical degrees of freedom in the system. In essence, the Lagrangian may serve to belie the subtle issues associated with counting degrees of freedom that are present in a physical system. 

These issues arise, for example, in electromagnetism, which displays U(1) gauge symmetry, and General Relativity (GR) which has gauge freedom due to diffeomorphism invariance. Cases in curved spacetime are of particular interest to those working in cosmology, as well as quantum gravity. For example, if one wishes to study the particle dynamics of a dark matter candidate when coupled to gravity, one may wish to work in a 3+1 formulation. It is important in such a scenario to be careful of the degrees of freedom, ensuring that the non-physical ones are discarded. This is where constraint reduction becomes useful.

Traditionally, constructing Hamiltonians from singular or constrained Lagrangians is either addressed on a case by case basis or using the Dirac~\cite{diraclectures} method. Dirac's technique~\citep{canadirac,1958RSPSA.246..326D} is a mechanism for solving the problem via phase space reduction. While the mechanism works, it can be challenging to work out.  Other approaches, known as symplectic techniques were developed, and the most enduring of these is that of Faddeev and Jackiw~\citep{1988PhRvL..60.1692F}, which relies on Darboux's theorem.

A useful feature of this technique is that it can be more straightforward than the Dirac approach, and it has been shown that the two methods achieve the same results in the most important, physically relevant cases~\citep{2003JPhA...36.1671R}. Another major difference between the Dirac approach and the Faddeev-Jackiw technique is that Dirac's immediately requires phase space reduction, while Faddeev-Jackiw enlarges the configuration space before breaking it into relevant and irrelevant pieces. As will be described in more detail below, this allows one to compute the Hamiltonian while postulating Poisson brackets, just as one might hope.  Of course, this is also achieved by the Dirac method~\citep{1997IJMPA..12..451G,1998IJMPA..13.3691G}, but sometimes with considerably greater effort.

Barcelos-Neto and Wotzasek~\citep{1992IJMPA...7.4981B} rework the Faddeev-Jackiw approach to avoid the challenge of needing to find the Darboux transformation. However, their approach does not capture all the essential details that are important for application to General Relativity and other theories with general covariance and gauge symmetry. Specifically, while Barcelos-Neto and Wotzasek address the application of the Faddeev-Jackiw technique to fields, certain subtleties are not addressed. As we will show in this paper, it is important to recognize and take advantage of the fact that the symplectic approach is fundamentally related to understanding the structure of the function space of the physical fields.

This paper successfully reformulates the Faddeev-Jackiw method for application to fields using the Ginzburg-Landau model in a curved background as an example. This model has enough complexity to provide a good basis for full general relativity (where the metric itself is dynamical), Yang-Mills theories, etc.  We restrict consideration to bosonic (commuting) fields; later work will consider fermionic (anti-commuting) fields. We begin in Sec.~\ref{FJtech} by describing the fundamentals of the Faddeev-Jackiw approach. In Sec.~\ref{BNWtech} we explicate the Barcelos-Neto-Wotzasek reformulation of this technique using the example of massive electromagnetism. In Sec.~\ref{new}, we introduce a new formulation of the technique, and we show how this formalism is applied to the example of the Ginzburg-Landau action in a curved background, a solution that has not previously appeared in the literature. We use metric signature $\{-,+,+,+\}$ throughout, and the Einstein summation convention with indices $i,j,k,l$ denoting spatial components and with indices $a,b,c$ denoting field components as explained below, unless otherwise noted.

\section{Faddeev-Jackiw Constraint Reduction}
\label{FJtech}

In this section, we give a brief overview of the original Faddeev and Jackiw paper following their original notation. The original Faddeev-Jackiw technique is based on the key insight that Darboux's theorem can be applied to a singular Lagrangian when it is written in first order, symplectic form. In this case, the Lagrangian takes the following form:
\begin{equation}\label{symlang}
L(\xi,\dot{\xi},t)dt=a_i d\xi^i -V(\xi) dt,
\end{equation}
where $\xi^i$ is a generalized coordinate that represents the position and momenta in phase space and the indices $i, j$ denote phase space components:

\begin{eqnarray}\label{sympvar}
\xi^i=p_i, i=1,...,n,\
\xi^i=q^{i-n}, i=n+1,...,2n.
\end{eqnarray}
Defining the symplectic matrix,
\begin{eqnarray}\label{sympform}
f_{ij}=\frac{\partial a_j}{\partial \xi^i}-\frac{\partial a_i}{\partial \xi^j},
\end{eqnarray}
we may rewrite the Euler-Lagrange equations:

\begin{eqnarray}\label{ELeqns}
f_{ij}\dot\xi^{j}=\frac{\partial V}{\partial \xi^i}.
\end{eqnarray}

If $f_{ij}$ is not singular, then it is straightforward to solve for the equations of motion to find

\begin{eqnarray}\label{invertEL}
\dot{\xi^i}=f^{-1}_{ij}\frac{\partial V}{\partial \xi^j},\
\{\xi^i , \xi^j\}=f^{-1}_{ij}.
\end{eqnarray}
In this case, the inverse of the symplectic form exists and its dimensions are $N=2n$. However, if it is singular, more work is necessary to find the proper equations of motion such that they formulate a true initial value problem. The presence of a singularity here indicates to us that the equations of motion contain more degrees of freedom than the actual number that exist in the system. In that case, the apparent dimensionality of the symplectic form is incorrect, and the right one will be $N=2n-N'$, where $N'$ is the number of Lagrange multipliers.

Faddeev and Jackiw recognized that one way to address this problem was to apply Darboux's theorem, which states that for any $a\equiv a_i d\xi^i$, it is always possible to change variables 
\begin{eqnarray}\label{darboux1}
\xi^i \rightarrow (p^j, q^k, z^l),\
j,k=1,...,n, l=1,...,N'
\end{eqnarray}
such that $a=p_i dq^i$ and
\begin{eqnarray}\label{darboux2}
L dt=p_i dq^i -\phi(p,q,z) dt, \frac{\partial \phi}{\partial z^l}=0.
\end{eqnarray}

After eliminating $z$ using the differential equations $\partial \phi /\partial z^l=0$, the Lagrangian becomes
\begin{eqnarray}
L = p_i \dot q^i - H(p,q) - \lambda_l \phi^{l}(p,q)
\end{eqnarray}
with $l^{th}$ constraint
\begin{eqnarray}
\phi^{l} = 0.
\end{eqnarray}
The Lagrangian can then be written in the form familiar from \ref{symlang}:
\begin{eqnarray}\label{newlagrange}
L dt = b_{i}(\eta) d\eta^i - W(\eta)dt
\end{eqnarray}
where $\eta$ is a new set of phase space variables that replace $\xi$.

We do not dwell here on the details but instead refer the reader to the original paper for more. Essentially, the procedure is repeated until (\ref{newlagrange}) takes the form of an unconstrained Lagrangian. It is then straightforward to apply the Legendre transformation and proceed to work in phase space. 

\section{The Barcelos-Neto-Wotzasek Implementation of the Faddeev-Jackiw Technique}\label{BNWtech}
As long as one knows the Darboux transformation, that is to find the transformation that gives the change of variables discussed in Sec.~\ref{FJtech}, the Faddeev-Jackiw technique is a readily accessible procedure. However, finding the transformation can be a challenge. Barcelos-Neto and Wotzasek~\cite{1992IJMPA...7.4981B} developed an algorithm to implement this technique that does not require foreknowledge of the Darboux transformation. This technique is also developed in Jackiw~\cite{notears}. 

We will reference the example of massive electromagnetism for the purposes of the discussion that follows. We start with the example of massive electromagnetism because it is simpler than the massless Ginzburg-Landau case since there is no gauge freedom but the Lagrangian is still singular.  The Lagrangian density is
\begin{equation}\label{LagMEM}
  {\cal L}=-\frac{1}{4}F^{\mu\nu}F_{\mu\nu}-\frac{1}{2}m^2A^\mu A_\mu+A_\mu J^\mu\ ,\ \ 
  F_{\mu\nu}\equiv\partial_\mu A_\nu-\partial_\nu A_\mu\ ,
\end{equation}
where $J^\mu$ is an external source and the Minkowski metric $\eta^{\mu\nu}$ is used to raise indices.  

Defining
\begin{equation}\label{Efield}
  E_i\equiv F_{i0}\ ,\ \ E^2\equiv E^iE_i\ ,\ \ F^2\equiv F^{ij}F_{ij}=2\vert\bm\nabla\times{\bm A}\vert^2\ , \ \ 
  A^2\equiv A^iA_i\ ,
\end{equation}
where $i=1, 2, 3$, the Lagrangian density may be written
\begin{equation}\label{LagMEM2}
  {\cal L}=\frac{1}{2}E^2-\frac{1}{4}F^2+\frac{1}{2}m^2\left(A_0^2-A^2\right)+A_\mu J^\mu\ .
\end{equation}
Given the fields $A_\mu$, the only nonzero canonical momenta are
\begin{equation}\label{piEM}
  \pi^i\equiv\frac{\partial\cal L}{\partial\dot A_i}=-E_i\ ,
\end{equation}
in terms of which the Hamiltonian becomes
\begin{eqnarray}\label{HamMEM}
  &&H[A_0,A_i,\pi^i]=\int d^3x\,{\cal H}\ ,\nonumber\\
  &&{\cal H}=\frac{1}{2}\pi^2+\frac{1}{4}F^2+\frac{1}{2}m^2\left(A^2-A_0^2\right)-A_iJ^i
    -A_0(\partial_i\pi^i+J^0)+\partial_i(\pi^iA_0)\ .\quad
\end{eqnarray}
Note that $\pi^2\equiv\pi^i\pi_i$.

The last term is a boundary term which makes no contributions to the equations of motion using the original variables, but which could contribute to them after the Hamiltonian reduction and therefore we retain it.  In general relativity such terms are related to conserved quantities that are due to Noether's theorem and which cannot be ignored~\citep{2000PhRvD..61h4027W}.

Finally, the action functional written in first-order form becomes
\begin{equation}\label{actionMEM}
  S[A_0,A_i,\pi^i]=\int d^4x\,(\pi^i\dot A_i)-\int dt\,H[A_0,A_i,\pi^i]\ .
\end{equation}
Our definition of the Hamiltonian omitted the term $\pi^0\dot A_0$ because this term is cancelled when the action is written in first-order form.

Varying the action with respect to the fields gives the equations of motion,
\begin{eqnarray}\label{EOMsMEM}
  A_0:&&0=\frac{\delta H}{\delta A_0}=-(\partial_i\pi^i+m^2A_0+J^0)\ \ \hbox{or}\ \ 
    {\bm\nabla}\cdot{\bm E}-m^2A_0=J^0\ ,\nonumber\\
  A_i:&&-\dot\pi^i=\frac{\delta H}{\delta A_i}=\partial_jF^{ij}+m^2A^i-J^i\ \ \hbox{or}\ \ 
    \partial_t{\bm E}-{\bm\nabla}\times({\bm\nabla}\times {\bm A})-m^2{\bm A}=-{\bm J}\ ,\nonumber\\
  \pi^i:&&\dot A_i=\frac{\delta H}{\delta\pi^i}=\pi_i+\partial_iA_0\ \ \hbox{or}\ \ 
    {\bm E}={\bm\nabla}A_0-\partial_t{\bm A}\ .
\end{eqnarray}

These are immediately recognizable as the Maxwell equations, modified by mass terms that explicitly break gauge-invariance (the potentials, not just the fields, appear in both the Poisson and Amp\`ere laws).  Indices have been raised or lowered in the second and third equations using $\delta_{ij}=\delta^{ij}$, which is valid in flat spacetime with the Minkowski metric.

The first step is to calculate the symplectic form. If it is invertible, the equations of motion can be solved as as an initial-value problem, then one may proceed as usual.  An ``initial-value problem'' is one in which the field values are arbitrarily specified on a constant-time surface and then advanced forward in time. A key reason we are interested in properly addressing the constraints is so that we can properly formulate an initial-value problem. In the case of a singular Lagrangian, we may be led to believe that there are more degrees of freedom than there actually are and that there is a well-defined initial-value problem when in fact there is not. It may appear as though we can arbitrarily specify $A_0$, $A_i$, and $\pi^i$ on a constant-time surface, when in fact only two of these variables are independent.  In other words, there is no time evolution equation for $A_0$.  Instead, it obeys an initial-value constraint, the first of equations (\ref{EOMsMEM}).  This is incompatible with Hamiltonian dynamics, which requires an initial-value problem as well as the ability to properly apply the Legendre transformation, which requires us to acknowledge and solve for the constraints.  To proceed with the Hamiltonian approach, we must find a good set of fields that obey an initial-value problem, and which are equivalent to equations (\ref{EOMsMEM}).

Having obtained an action for all the fields, the next step is to get the equations of motion in symplectic form as done by Fadeev and Jackiw. In the Hamiltonian picture, we use generalized coordinates and momenta that are combined into a set of phase space field variables that we designate $\xi_a$.  In the current example, $\xi_a\in\{A_0,A_i,\pi^i\}$.  The action in Hamiltonian form consists of a symplectic part called $S_{\rm sym}[\xi]$ that depends linearly on time derivatives of these quantities, and a part that is independent of time derivatives:
\begin{equation}\label{actionsplit}
  S[\xi]=S_{\rm sym}[\xi]-\int dt\,H[\xi]\ .
\end{equation}
Requiring the full action to be stationary under variation of the fields gives Hamilton's equations
\begin{equation}\label{Hamfield}
  \frac{\delta S_{\rm sym}}{\delta\xi_a}({\bm x},t)\equiv\int d^3y\,f_{\xi_a\xi_b}({\bm x},{\bm y})\dot\xi_b({\bm y},t)
    -\dot S_{\xi_a}({\bm x},t)=\frac{\delta H}{\delta\xi_a}({\bm x},t)\ ,
\end{equation}
where $\dot{S_{\xi_a}}$, if nonzero, is an external source appearing in $S_{\rm sym}$ as a term $\int d^4x\,S_{\xi_b}\dot\xi^b$. It will become clear in Sec.~\ref{new} that if the correct fields are used, terms like this one are unnecessary. Einstein summation is used on the repeated index $\xi_b$.  The antisymmetric matrix of fields $f_{\xi_a\xi_b}({\bm x},{\bm y})$, called the symplectic matrix, is found by varying $S_{\rm sym}$ and following the definition given by the first term in equation (\ref{Hamfield}); in general, this will require introducing a three-dimensional Dirac delta function, as we demonstrate below.

For a well-posed Hamiltonian problem, $f_{\xi_a\xi_b}({\bm x},{\bm y})$ is non-singular and can be inverted to give uniquely a matrix of fields $\tilde f_{\xi_a\xi_b}({\bm x},{\bm y})$ which obeys the following equations
\begin{equation}\label{ffabinv}
  \int d^3z\,\tilde f_{\xi_a\xi_c}({\bm x},{\bm z})f_{\xi_c\xi_b}({\bm z},{\bm y})=
  \int d^3z\,f_{\xi_a\xi_c}({\bm x},{\bm z})\tilde f_{\xi_c\xi_b}({\bm z},{\bm y})=\delta_{\xi_a\xi_b}\delta^3({\bm x}-{\bm y})\ .
\end{equation}
If $\tilde f$ exists, then equation (\ref{Hamfield}) can be solved to give the time evolution of the fields,
\begin{equation}\label{xievol}
  \dot\xi_a({\bm x},t)=\int d^3y\,\tilde f_{\xi_a\xi_b}({\bm x},{\bm y})\left[\frac{\delta H}{\delta\xi_b}({\bm y},t)
    +\dot S_{\xi_b}({\bm y},t)\right]\ ,
\end{equation}
from which the solution of the initial-value problem proceeds.  These time evolution equations are equivalent to, and have the same solutions, as the Euler-Lagrange equations arising from the original action.  Note that $\tilde f$ is antisymmetric just like $f$, i.e.
\begin{equation}\label{fantisym}
  \tilde f_{\xi_a\xi_b}({\bm x},{\bm y})=-\tilde f_{\xi_b\xi_a}({\bm y},{\bm x})\ .
\end{equation}

When the Lagrangian is singular, $f_{\xi_a\xi_b}$ cannot be inverted.  For example, with action (\ref{actionMEM}),  the only non-zero elements $f_{\xi_a\xi_b}$ are
\begin{equation}\label{fab0}
  f_{\pi^jA_i}({\bm x},{\bm y})=-f_{A_i\pi^j}({\bm y},{\bm x})=\delta^i_{\ j}\,\delta^3({\bm x}-{\bm y})\ .
\end{equation}
All elements $f_{\xi_a\xi_b}({\bm x},{\bm y})$ with $\xi_a=A_0$ or $\xi_b=A_0$ vanish in our example, implying that $f_{\xi_a\xi_b}$ is singular and $\tilde f_{\xi_a\xi_b}$ does not exist.  There is a zero-mode of $f_{\xi_a\xi_b}$, i.e., an eigenvector with zero eigenvalue.  The Faddeev-Jackiw method introduces additional field variables so that, in the new fields, the resulting $f_{\xi_a\xi_b}$ is non-singular.  This method proceeds by adding constraints to the Lagrangian, as we now show for the example of massive electromagnetism.

Following Barcelos-Neto and Wotzasek, we call the constraint corresponding to the zero-mode $\Omega\equiv-\delta H/\delta A_0=0$, enforce it using a Lagrange multiplier $\dot\lambda$ so that $\dot\Omega=0$ becomes an equation of motion, and define a new Hamiltonian system as follows:
\begin{equation}\label{Action1}
  S^1_{\rm sym}[\xi^1]\equiv\int d^4x\,(\pi^i\dot A_i+\dot\lambda\Omega)\ ,\ \ 
  \Omega=-\frac{\delta H}{\delta A_0}=\partial_i\pi^i+m^2A_0+J^0\ ,
\end{equation}
with Hamiltonian
\begin{equation}\label{Ham1}
  H_1[\xi^1]=\int d^3x\,{\cal H}_1\ ,\ \ 
  {\cal H}_1={\cal H}\vert_{\Omega=0}\ .
\end{equation}
The set of fields has been increased by one, $\xi^1_a\in\{A_0,\lambda,A_i,\pi^j\}$, where the superscript 1 and the similar subscript on $H_1$ denote the first correction to the original fields and Hamiltonian.  (The process may be iterated, leading to a second correction or more.)  As we will see later, the ambiguity in how the constraint $\Omega=0$ is applied to convert $H$ to $H_1$ disappears when we seek out all the constraints before imposing the constraints on the Hamiltonian. This subtlety is not addressed in the BNW formulation.  

We require that $\delta S^1_{\rm sym}/\delta\xi^1_a=\delta H_1/\delta\xi^1_a$ yield the same equations of motion as $\delta S_{\rm sym}/\delta\xi_a=\delta H/\delta\xi_a$.  Importantly, ${\cal H}_1$ must not depend on the Lagrange multiplier $\lambda$; the only place where $\lambda$ enters is in the symplectic action $S^1_{\rm sym}$.  For completeness, note that in this example one of the new terms, $\dot\lambda J^0$, involves the external source $S_\lambda=J^0$ instead of one of the electromagnetic fields $\{A_0,A_i,\pi^j\}$.

Given an action in first-order form, we first evaluate the new $f^1_{\xi_a\xi_b}({\bm x},{\bm y})$ (whose subscript is written $\xi_a\xi_b$ instead of $\xi^1_a\xi^1_b$ to avoid clutter) and determine whether it is singular.  If $f^1_{\xi_a\xi_b}$ is non-singular, invert it; otherwise, find a zero-mode and add it as a constraint with a new Lagrange multiplier, enforcing the old and new constraints in the Hamiltonian.  The equations of motion must also be checked, which is essential to determining $H_1$.

Using the definition in equation (\ref{Hamfield}), $f^1_{\xi_a\xi_b}({\bm x},{\bm y})$ has nonzero elements
\begin{eqnarray}\label{fab1}
  &&f^1_{\pi^jA_i}({\bm y},{\bm x})=-f^1_{A_i\pi^j}({\bm x},{\bm y})=\delta^i_{\ j}\,\delta^3({\bm x}-{\bm y})\ ,\nonumber\\
  &&f^1_{A_0\lambda}({\bm y},{\bm x})=-f^1_{\lambda A_0}({\bm x},{\bm y})=m^2\,\delta^3({\bm x}-{\bm y})\ ,\nonumber\\
  &&f^1_{\pi^i\lambda}({\bm y},{\bm x})=-f^1_{\lambda\pi^i}({\bm x},{\bm y})
    =\frac{\partial}{\partial x^i}\delta^3({\bm x}-{\bm y})=-\frac{\partial}{\partial y^i}\delta^3({\bm x}-{\bm y})\ .
\end{eqnarray}
The derivative of the delta function comes from integrating by parts $\dot\lambda\partial_i\pi^i$ in the symplectic action (\ref{actionMEM}), and may be checked by substitution into the second of equations (\ref{Hamfield}): 
\begin{equation}\label{checkfab1}
  \frac{\delta S^1_{\rm sym}}{\delta\pi^i}=\dot A_i-\partial_i\dot\lambda\ ,
\end{equation}
while from the second of equations (\ref{Hamfield}) with equations (\ref{fab1}),
\begin{eqnarray}\label{check2fab1}
  \frac{\delta S^1_{\rm sym}}{\delta\pi^i}&=&\int d^3y\left[f_{\pi^iA_j}({\bm x},{\bm y})\dot A_j({\bm y},t)
    +f^1_{\pi^i\lambda}({\bm x},{\bm y})\dot\lambda({\bm y},t)\right]\nonumber\\
  &=&\int d^3y\left[\delta^j_{\ i}\delta^3({\bm x}-{\bm y})\dot A_j({\bm y},t)-\frac{\partial}{\partial x^i}
    \delta^3({\bm x}-{\bm y})\dot\lambda({\bm y},t)\right]\nonumber\\
  &=&\dot A_i({\bm x},t)-\frac{\partial}{\partial x^i}\dot\lambda({\bm x},t)\ ,
\end{eqnarray}
in agreement with (\ref{checkfab1}).

Now, it happens that $f^1_{\xi_a\xi_b}$ is non-singular, with inverse
\begin{eqnarray}\label{finvab1}
  &&\tilde f^{\,1}_{A_i\pi^j}({\bm x},{\bm y})=-\tilde f^{\,1}_{\pi^jA_i}({\bm y},{\bm x})
    =\delta^i_{\ j}\,\delta^3({\bm x}-{\bm y})\ ,\nonumber\\
  &&\tilde f^{\,1}_{\lambda A_0}({\bm x},{\bm y})=-\tilde f^{\,1}_{A_0\lambda}({\bm y},{\bm x})
    =\frac{1}{m^2}\,\delta^3({\bm x}-{\bm y})\ ,\nonumber\\
  &&\tilde f^{\,1}_{A_0A_i}({\bm x},{\bm y})=-\tilde f^{\,1}_{A_iA_0}({\bm y},{\bm x})
    =\frac{1}{m^2}\frac{\partial}{\partial x^i}\delta^3({\bm x}-{\bm y})
    =-\frac{1}{m^2}\frac{\partial}{\partial y^i}\delta^3({\bm x}-{\bm y})\ .
\end{eqnarray}
The proof is to substitute equations (\ref{fab1}) and (\ref{finvab1}) into (\ref{ffabinv}).

Ideally, the next step is to reduce the Hamiltonian to canonical pairs. However, there is a fundamental flaw in this algorithm. While the case of massive electromagnetism is simple enough that problems do not arise, application of this algorithm to a more complicated action with multiple constraints will cause problems if the constraints are enforced too early. This does not preclude one from using the constraint; we simply propose that they should not be eliminated. This is discussed further in the next section. 
\section{A New Approach to Faddeev and Jackiw}
\label{new}
We now introduce our new approach in order to solve the problem of the singular Lagrangian for massless electromagnetism. A notable divergence from section~\ref{BNWtech} is the absence of an explicit calculation of the symplectic matrix, although effectively we are computing some of its components.  In other words, we will not use the $f_{ab}$ matrix approach of Barcelos-Neto and Wotzasek to construct a non-singular Lagrangian.  Instead, we work directly with a functional approach.

The approach is inspired by earlier work of Maskawa and Nakajima~\cite{1976PThPh..56.1295M}, Faddeev and Jackiw, and Barcelos-Neto and Wotzasek, although there are notable differences.  The method is simpler and incorporates several new ideas for gauge theories in curved spacetime.  The current work is based on four general principles:
\begin{enumerate}
\item The fields appearing in the action can be transformed without changing the physics, provided that the new fields span the entire function space (so, for example, the gradient of a scalar field does not span the same space as an unconstrained vector field).
\item A Hamiltonian formulation means that every field appearing in the action obeys canonical Poisson brackets and has time evolution generated by the Hamiltonian.
\item Symmetries of both the full action and the symplectic action lead to constraints.  Constraints are classical equations of motion without time evolution.  Constraints must be converted to canonical fields consistent with the second principle.
\item Constraints are given time evolution by adding them to the action with Lagrange multipliers.
\end{enumerate}

The Hamiltonian reduction has three steps: identification of constraints, construction of a good Hamiltonian for all fields including the constraints and Lagrange multipliers, and finally the elimination of the constraints.  Following Maskawa and Nakajima, this is done by partitioning phase space into a product of unconstrained (dynamical) and constrained subspaces. This division of the phase space is a critical insight for comprehending the holistic problem of the symplectic structures of systems described by singular Lagrangians. 

There are three types of constraints, distinguished only by their origin: primary constraints, defined as zero modes of the symplectic action following from the original Legendre transformation (i.e. $H^0$); gauge constraints, which are zero modes of the total action; and secondary constraints, which follow from time derivatives of other constraints. It does not matter how the constraints are found, only that enough are found to achieve a complete Hamiltonian formulation. The use of classical equations of motion, symmetries of the action, and Faddeev-Jackiw iteration are equally valid methods for finding constraints. 

Each constraint adds a new field to the action, and constraints might depend on ones found earlier. Consequently the order in which constraints are introduced can affect the search for new constraints.  (This happens when there are second class constraints in Dirac's terminology.)  All constraints have been found when all gauge constraints are included in the action and the symplectic action is nonsingular, i.e. when all equations of motion are time evolution equations. For every constraint there is a corresponding Lagrange multiplier (a time derivative of a new field).

The goal of Hamiltonian reduction is to find enough constraints and Lagrange multipliers to span the original function space (and, in general, a larger space), and the enlarged function space is a symplectic manifold. This goal is achieved when all fields are present in a symplectic action of canonical form. 

\subsection{Ginzburg-Landau Action}\label{GLA}
The classical field theory for a complex scalar field representing spinless particles of charge $q$ and mass $m$ is known as the Ginzburg-Landau (GL) model~\cite{1957JPCS....2..199A}. This is also known as the Abelian Higgs action~\cite{2006stmo.book.....B}, but we will use the Ginzburg-Landau interpretation of it. Essentially, the GL action describes a charged fluid that is coupled to the electromagnetic field. In this section, we work in curved spacetime, so the scalar fields and electromagnetism are also coupled to gravity. The utility of this model lies in its gauge and diffeomorphism invariances, which are fundamental properties of the standard model of particle physics and general relativity, respectively. Although it was originally developed as an effective field theory with application to superconductivity~\cite{2004RvMP...76..981G}, the GL model is a useful toy model for exploring the issues that come up in the study of the standard model on a background that is not Minkowski. 

In covariant form, the Ginzburg-Landau Lagrangian density is
\begin{equation}\label{Lagrangian1}
  {\cal L}=-\frac{1}{4}\sqrt{-g}\,g^{\mu\kappa}g^{\nu\lambda}
    F_{\mu\nu}F_{\kappa\lambda}-\frac{1}{2}\sqrt{-g}\left[g^{\mu\nu}
    (D_\mu\phi)^\ast(D_\nu\phi)-\mu^2|\phi|^2+\frac{\lambda}{2}|\phi|^4\right]
\end{equation}
where $F_{\mu\nu}\equiv\partial_\mu A_\nu-\partial_\nu A_\mu$, $\phi$ is a complex scalar field,
\begin{equation}\label{phidef}
  \phi=\phi_1+i\phi_2\ ,
\end{equation}
and $D_\mu$ is the gauge covariant derivative,
\begin{equation}\label{gcder}
  \ D_\mu\equiv\nabla_\mu-iqA_\mu\ .
\end{equation}
This gives rise to the Euler-Lagrange equations
\begin{eqnarray}\label{ELcovar}
  g^{\mu\nu}D_\mu D_\nu\phi&=&(-\mu^2+\lambda\phi^\ast\phi)\phi\ ,\nonumber\\
  \nabla_\nu F^{\mu\nu}&=&\frac{1}{2}qg^{\mu\nu}[i\phi(D_\nu\phi)^\ast-i\phi^\ast(D_\nu\phi)]
  \equiv J^\mu\ .
\end{eqnarray}

It is clear that this action will show U(1) gauge symmetry, and coupling between the scalar field and the electromagnetic field is introduced by the gauge covariant derivative. The appearance of a generic metric couples the fields to gravity. The first two terms of the action are familiar kinetic and quadratic potential terms, respectively. The quartic term represents self-interaction that lead to spontaneous symmetry breaking. When working with a scalar field, $\nabla_\mu$ reduces to a simple partial derivative. 

We perform a 3+1 split by writing the metric in the ADM form,
\begin{equation}\label{metric}
  g_{00}=-\alpha^2+\beta^2\ ,\ \ g_{0i}=\beta_i\ ,\ \ g_{ij}=\gamma_{ij}\ ,
\end{equation}
with inverse metric
\begin{equation}\label{invmetric}
  g^{00}=-\frac{1}{\alpha^2}\ ,\ \ g^{0i}=\frac{\beta^i}{\alpha^2}\ ,\ \ g^{ij}=\gamma^{ij}-\frac{\beta^i\beta^j}{\alpha^2}\ ,
\end{equation}
where $\gamma^{ij}$ is the 3-dimensional inverse of $\gamma_{ij}$ and is used to raise spatial indices,
\begin{equation}\label{spatial}
  \gamma^{ik}\gamma_{kj}=\delta^i_{\ \,j}\ ,\ \ \beta^i\equiv\gamma^{ij}\beta_j\ ,\ \ \beta^2\equiv\beta^i\beta_i\ .
\end{equation}
With these definitions, $\sqrt{-g}=\alpha\sqrt{\gamma}$ where $\gamma={\rm det}(\gamma_{ij})$.

Using the 3+1 split, the Lagrangian becomes
\begin{eqnarray}\label{Lagrangian2}
  {\cal L}&=&\frac{\sqrt{\gamma}}{2\alpha}\gamma^{ij}(F_{i0}-\beta^kF_{ik})
    (F_{j0}-\beta^lF_{jl})-\frac{1}{4}\alpha\sqrt{\gamma}\gamma^{ik}\gamma^{jl}
     F_{ij}F_{kl}\nonumber\\
  &&+\frac{\sqrt{\gamma}}{2\alpha}\vert D_0\phi-\beta^iD_i\phi\vert^2
    -\frac{\alpha\sqrt{\gamma}}{2}\left[\gamma^{ij}
    (D_i\phi)^\ast(D_j\phi)-\mu^2|\phi|^2+\frac{\lambda}{2}|\phi|^4\right]\ .
\end{eqnarray}
Varying the action gives
\begin{eqnarray}\label{VaryAction1}
  \frac{\delta S}{\delta A_0}&=&\partial_i\pi^i-\frac{1}{2}q(i\phi\pi^\ast-i\phi^\ast\pi)
    \ ,\nonumber\\
  \frac{\delta S}{\delta A_i}&=&-\dot\pi^i-\partial_j\left(\alpha\sqrt{\gamma}F^{ij}
    +\beta^i\pi^j-\beta^j\pi^i\right)+\frac{1}{2}q\beta^i(i\phi\pi^\ast-i\phi^\ast\pi)
    \nonumber\\
    &&+\frac{1}{2}q\alpha\sqrt{\gamma}\gamma^{ij}\left(i\phi\partial_j\phi^\ast
    -i\phi^\ast\partial_j\phi\right)-q^2\vert\phi\vert^2\alpha\sqrt{\gamma}\gamma^{ij}A_j
    \ ,\nonumber\\
  \left(\frac{\delta S}{\delta\phi}\right)^\ast&=&-\dot\pi+iq(A_0-\beta^iA_i)\pi
    +\partial_i\left(\beta^i\pi+\alpha\sqrt{\gamma}\gamma^{ij}D_j\phi\right)
    -iq\alpha\sqrt{\gamma}\gamma^{ij}A_iD_j\phi\nonumber\\
    &&+\alpha\sqrt{\gamma}\left(\mu^2-\lambda\vert\phi\vert^2\right)\phi\ ,
\end{eqnarray}
where we have defined
\begin{eqnarray}\label{pidef}
  \pi^i&\equiv&\frac{\partial{\cal L}}{\partial\dot A_i}
    =\frac{\sqrt{\gamma}}{\alpha}\gamma^{ij}\left(\dot A_j-\partial_j A_0+\beta^kF_{jk}
    \right)\ ,\nonumber\\
  \pi&\equiv&\left(\frac{\partial{\cal L}}{\partial\dot\phi}\right)^\ast
    =\frac{\sqrt{\gamma}}{\alpha}\left(D_0\phi-\beta^iD_i\phi\right)=\pi_1+i\pi_2
    \ ,\nonumber\\
  \pi_1&=&\frac{\sqrt{\gamma}}{\alpha}\left(\dot\phi_1-\beta^i\partial_i\phi_1
    +q\tilde A\phi_2\right)\ ,\nonumber\\
  \pi_2&=&\frac{\sqrt{\gamma}}{\alpha}\left(\dot\phi_2-\beta^i\partial_i\phi_2
    -q\tilde A\phi_1\right)\ ,
\end{eqnarray}
with
\begin{equation}\label{Atilde}
  \tilde A\equiv A_0-\beta^iA_i\ 
\end{equation}
and complex conjugates are found in the usual manner.

Using the definition of the canonical momenta and writing out separately the real and imaginary parts of the fields, the classical equations of motion are
\begin{eqnarray}\label{ELeqs}
  0&=&-\partial_i\pi^i+q(\phi_1\pi_2-\phi_2\pi_1)\ ,\nonumber\\
  \dot A_i&=&\frac{\alpha}{\sqrt{\gamma}}\gamma_{ij}\pi^j+\partial_iA_0-\beta^kF_{ik}
    \ ,\nonumber\\
  -\dot\pi^i&=&\partial_j\left(\alpha\sqrt{\gamma}F^{ij}+\beta^i\pi^j-\beta^j\pi^i\right)
    -q\beta^i(\phi_1\pi_2-\phi_2\pi_1)\nonumber\\
    &&-q\alpha\sqrt{\gamma}\gamma^{ij}\left(\phi_1\partial_j\phi_2-\phi_2\partial_j\phi_1
    \right)+q^2\left(\phi_1^2+\phi_2^2\right)\alpha\sqrt{\gamma}\gamma^{ij}A_j
    \ ,\nonumber\\
  \dot\phi_1&=&\frac{\alpha}{\sqrt{\gamma}}\pi_1+\beta^i\partial_i\phi_1
    -q\tilde A\phi_2\ ,\nonumber\\
  \dot\phi_2&=&\frac{\alpha}{\sqrt{\gamma}}\pi_2+\beta^i\partial_i\phi_2
    +q\tilde A\phi_1\ ,\nonumber\\
  -\dot\pi_1&=&q\tilde A\pi_2-\partial_i\left[\beta^i\pi_1+\alpha\sqrt{\gamma}\gamma^{ij}
    (\partial_j\phi_1+qA_j\phi_2)\right]-q\alpha\sqrt{\gamma}\gamma^{ij}A_i
    \left(\partial_j\phi_2-qA_j\phi_1\right)\nonumber\\
    &&+\alpha\sqrt{\gamma}\left[-\mu^2+\lambda\left(\phi_1^2+\phi_2^2\right)\right]
    \phi_1\ ,\nonumber\\
  -\dot\pi_2&=&-q\tilde A\pi_1-\partial_i\left[\beta^i\pi_2+\alpha\sqrt{\gamma}\gamma^{ij}
    (\partial_j\phi_2-qA_j\phi_1)\right]+q\alpha\sqrt{\gamma}\gamma^{ij}A_i
    \left(\partial_j\phi_1+qA_j\phi_2\right)\nonumber\\
    &&+\alpha\sqrt{\gamma}\left[-\mu^2+\lambda\left(\phi_1^2+\phi_2^2\right)\right]
    \phi_2\ .
\end{eqnarray}

Clearly, the first equation is an initial-value constraint, not an evolution equation, indicating a singular Lagrangian.  What is not so obvious is that the equations of motion, and the action from which they are derived, are invariant under the gauge transformation
\begin{equation}\label{gaugex}
  A_\mu\to A_\mu-\partial_\mu\chi_g\ ,\ \ \phi\to\exp(-iq\chi_g)\phi\ ,\ \ 
  \pi\to\exp(-iq\chi_g)\pi\ .
\end{equation}
As a result, the Lagrangian contains both too few degrees of freedom (there is no momentum corresponding to $A_0$) and too many (there is a gauge mode $\chi_g$).  As we will see in the next section, these features complicate the Hamiltonian formulation.

Before moving on, it is worth noting that equations (\ref{ELeqs}) provide a gauge-invariant way to provide the photon with mass.  In the ground state, where fields are constant, $\vert\phi\vert^2=\mu^2/\lambda$ on account of the scalar field potential having its minimum at nonzero field values.  As a result, the equation for $-\dot\pi^i$ has a photon mass term.  Consequently, the symmetry breaking that occurs when the scalar field is nonzero gives rise to photon mass.  This is an Abelian version of the Higgs mechanism, and is well known in condensed matter physics as a model for superconductivity.  Although the photon mass term proportional to $q^2(\phi_1^2+\phi_2^2)$ is not gauge-invariant, the combination of this term and the one proportional to $\phi_1\partial_j\phi_2-\phi_2\partial_j\phi_1$ is gauge-invariant.

\subsection{Constraints in the first-order formulation}\label{GLA2}
The Lagrangian (\ref{Lagrangian1}) is nonlinear in the time derivatives of the fields.  The Hamiltonian formulation replaces time derivatives of the fields with canonical momenta, following the first general principle stated at the beginning of Sec.~\ref{new}. The construction of a satisfactory Hamiltonian description means finding the right set of fields to give an appropriate initial-value formulation with Poisson brackets.  Constraints complicate this construction.

The first step towards finding an initial-value formulation is to construct a first-order (in time derivatives) action by replacing the time derivatives of the fields with the momenta defined in equations  (\ref{pidef}), yielding
\begin{equation}\label{Action0}
  S[A_0,A_i,\pi^i,\phi_1,\phi_2,\pi_1,\pi_2]=\int d^4x(\pi^i\dot A_i+\pi_1\dot\phi_1
    +\pi_2\dot\phi_2)-\int dt\,H^0[A_0,A_i,\pi^i,\phi_1,\phi_2,\pi_1,\pi_2]\ ,
\end{equation}
where
\begin{eqnarray}\label{H0def}
  H^0&=&\int d^3x\Biggl\{\frac{\alpha}{2\sqrt{\gamma}}\gamma_{ij}\pi^i\pi^j
    +\pi^i\left(\partial_iA_0-\beta^kF_{ik}\right)+\frac{1}{4}\alpha\sqrt{\gamma}
    \gamma^{ik}\gamma^{jl}F_{ij}F_{kl}+\frac{\alpha}{2\sqrt{\gamma}}
    \left(\pi_1^2+\pi_2^2\right)\nonumber\\
  &&+\beta^i\left(\pi_1\partial_i\phi_1+\pi_2\partial_i\phi_2\right)+q\tilde A(\phi_1\pi_2
    -\phi_2\pi_1)+\frac{\alpha\sqrt{\gamma}}{2}\left[-\mu^2\left(\phi_1^2+\phi_2^2\right)
    +\frac{1}{2}\lambda\left(\phi_1^2+\phi_2^2\right)^2\right]\nonumber\\
  &&+\frac{\alpha\sqrt{\gamma}}{2}\gamma^{ij}\left[(\partial_i\phi_1+qA_i\phi_2)    
    (\partial_j\phi_1+qA_j\phi_2)+(\partial_i\phi_2-qA_i\phi_1)(\partial_j\phi_2-qA_j\phi_1)
    \right]\Biggr\}
\end{eqnarray}
The first part of (\ref{Action0}) is called the symplectic part, and as written it has separated the fields $(A_i,\phi_1,\phi_2)$ and the momenta $(\pi^i,\pi_1,\pi_2)$ in perfect symplectic form except that $A_0$ and an associated momentum are missing.  Varying the action with respect to all of the fields gives exactly equations (\ref{ELeqs}).  In this sense, equations (\ref{Action0})--(\ref{H0def}) provide a good system.  However, because there is no momentum field corresponding to $A_0$, these equations do not provide an initial value problem nor do they provide good Poisson brackets.  To fix this, we have to add new fields to the action, including constraints and Lagrange multipliers.

Although we will call the function $H^0$ and its refinements the Hamiltonian, until the full action has the appropriate form, we will not have a fully satisfactory Hamiltonian system.  Obtaining one requires first finding the constraints, which is done in this section, and then finding Poisson brackets, which is done in the next section.  A constraint is simply an equation of motion without time derivatives of the fields.

The first step is to find as many constraints as we can solely in terms of the fields appearing in $H^0$.  Constraints may be zero modes of the symplectic action, zero modes of the total action, gauge constraints or time derivatives of any of the other constraints (re-expressed in terms of the fields appearing in $H^0$).  Note that the finding of constraints can proceed without the introduction of Lagrange multipliers. This is one way our technique departs from the original Faddeev-Jackiw algorithm. The Darboux transformation is not necessary and at this stage, neither are the Lagrange multipliers. 

The first and most obvious constraint is the first of equations (\ref{ELeqs})
\begin{equation}\label{Omega1}
  \Omega_1\equiv\partial_i\pi^i+\rho\ ,\ \ \rho\equiv q(\phi_2\pi_1-\phi_1\pi_2)\ .
\end{equation}
This constraint is called a primary constraint because it comes directly from Hamilton's equations using the original Hamiltonian.  It is a zero-mode of the symplectic part of the action with
\begin{equation}\label{varyA0}
  \delta A_0=\chi_1\ ,\ \ \delta A_i=\delta\pi^i=\delta\phi_1=\delta\phi_2=\delta\pi_1
    =\delta\pi_2=0\ .
\end{equation}
That is, the field variation $\chi_1({\bm x},t)$ leads to a vanishing first-order change of the symplectic action in (\ref{Action0}) because $A_0$ does not appear in the symplectic part.  Consequently the classical equation of motion associated with $\chi_1$ is
\begin{equation}\label{Constraint1}
  -\Omega_1=\frac{\delta H_0}{\delta\chi_1}=\frac{\delta H_0}{\delta A_0}
\end{equation}
in agreement with equation (\ref{Omega1}).

It is very important in what follows that we regard $\Omega_1({\bm x},t)$, like any constraint, as a field whose value depends on the other fields in the action, and that we distinguish the definition of this field from the imposition of the constraint $\Omega_1({\bm x},t)=0$.  To find a good initial-value formulation, we promote all constraints to fields on equal footing with every other field in the action.  We do not assume at the outset that the constraints will vanish.  Once we have found a good Hamiltonian, its equations of motion will dynamically enforce the vanishing of all constraints.

One might try to take a short-cut by immediately applying the constraint $\Omega_1=0$ directly in the action (\ref{Action0}).  This will simplify the Hamiltonian; $A_0$ drops out completely.  However, the equations of motion resulting from this amputated action are incorrect.  The Hamiltonian can only be reduced after all of the constraints have been found and new phase space variables that partition phase space into distinct dynamical and constrained subspaces have been chosen.  Only at that point will we be able to eliminate the constraints.  In general, there is no shortcut to the process. Certain simple cases such as massive electromagnetism turn out to be solvable taking a shortcut, but that is an exception rather than the rule.

Another obvious constraint follows from gauge-fixing, which eliminates the zero-mode $\chi_g$ of the full action shown by equation (\ref{gaugex}).  The simplest way to achieve this is by using Coulomb gauge, which is defined by the constraint
\begin{equation}\label{Omegag}
  \Omega_g\equiv\sqrt{\gamma}\nabla^iA_i=\partial_i\left(\sqrt{\gamma}\gamma^{ij}A_j
    \right)\ .
\end{equation}
As with $\Omega_1$, we define $\Omega_g$ in terms of the primary fields in the action, and regard it as a new field whose dynamics will later follow from a complete Hamiltonian.

Having found two constraints, we search for two more by taking time derivatives and using the classical equations of motion.  Differentiating equations (\ref{Omega1}) and (\ref{Omegag}) and using equations (\ref{ELeqs}) gives $\dot\Omega_1\equiv0$ and
\begin{equation}\label{Omegas}
  \dot\Omega_g=\Omega_s\equiv\partial_i\left[\alpha\pi^i+\sqrt{\gamma}
    \gamma^{ij}\left(\partial_jA_0-\beta^kF_{jk}\right)+A_j\partial_t\left(\sqrt{\gamma}
    \gamma^{ij}\right)\right]\ .
\end{equation}
The primary constraint $\Omega_1$ produces no more constraints, while the gauge constraint generates a secondary constraint $\Omega_s$, a combination of the canonical fields without time derivatives.  (Note that the spatial metric is an external field whose time derivative is allowed in an initial-value constraint.)

Have we found all of the constraints?  As stated in the introduction, we expect three kinds of constraints: primary constraints of the original action, gauge constraints, and time derivatives of the other constraints.  This suggests there might be another constraint arising from $\dot\Omega_s$.  However, the derivative leads us to $\dot A_0$, for which there is no equation of motion, so we cannot obtain another constraint.

The real test comes when we add the constraints to the action with Lagrange multiplier fields, evaluate the equations of motion, and see whether there is a time evolution equation for every field that reproduces the original set (\ref{ELeqs}).  This is done in the next section.

\subsection{Calculating the Poisson Brackets}\label{GLA3}

If we have found a complete set of constraints for the action, what remains is to put the action in the standard symplectic form
\begin{equation}\label{symplectic}
  S=\int d^4x\,\sum_a\pi_a\dot\phi_a-\int dt\,H[\phi_a,\pi_a]
\end{equation}
with a set of independent fields $(\phi_a,\pi_a)$ that spans the function space of the original fields plus constraints and Lagrange multipliers.  Once this is achieved the constraints will be enforced by the dynamics, i.e. the equations of motion, and can be dropped from the function space along with their Lagrange multipliers by partitioning the function space into a product of constrained and unconstrained spaces~\cite{1976PThPh..56.1295M}.  Achieving this requires three steps: putting the action in symplectic form with the constraints included, finding the Poisson brackets, and then eliminating the constraints.  The first step will be achievable if and only if we have found all of the constraints.  If we have not, the deviation from symplectic form will guide us to finding any remaining constraints.

Constraint fields are added to the symplectic part of the action with Lagrange multipliers:\begin{equation}\label{Symplectic1}
  I^1=\int d^4x\left(\pi^i\dot A_i+\pi_1\dot\phi_1+\pi_2\dot\phi_2+\dot\lambda_1\Omega_1
    +\dot\lambda_g\Omega_g+\dot\lambda_s\Omega_s\right)
\end{equation}
where
\begin{equation}\label{Action1}
  S=I^1-\int dt\,H^0[A_0,A_i,\pi^i,\phi_1,\phi_2,\pi_1,\pi_2]\ .
\end{equation}
The Lagrange multipliers fields $\dot\lambda_1$, $\dot\lambda_g$, and $\dot\lambda_s$ must be introduced with time derivatives in order to give dynamics in the Hamiltonian formulation.  Their purpose is to promote the constraints to dynamical fields obeying time evolution equations.

At first glance equations (\ref{Symplectic1}) and (\ref{Action1}) appear to have succeeded in the first step of putting the action into symplectic form with the constraints (and Lagrange multipliers) included.  However, there is no momentum field corresponding to $A_0$, and the constraints are not independent fields but are abbreviations for combinations of the other fields $(A_0,A_i,\pi^i,\phi_1,\phi_2,\pi_1,\pi_2)$ defined by equations (\ref{Omega1}), (\ref{Omegag}) and (\ref{Omegas}).  Thus the fields in the argument lists of (\ref{Symplectic1}) and (\ref{Action1}) do not properly span the function space; for example, $\Omega_g$ and $A_i$ cannot both be chosen independently, which is a requirement for the fields in the argument list of the action.

These considerations may lead one to choose the set of variables appearing in the action as
\begin{equation}\label{field1}
  \xi_a\in\left\{A_0,A_i,\pi^i,\phi_1,\phi_2,\pi_1,\pi_2,\lambda_1,\lambda_g,\lambda_s\right\}\ .
\end{equation}
All of the fields $\xi_a$ appear in the symplectic action (\ref{Symplectic1}) (although not in the standard symplectic form) and they fully span the Hilbert space because $\xi_a$ includes the 7 original fields used to describe the system plus three fields associated with the constraints (the Lagrange multipliers).  It can be shown that equations (\ref{Symplectic1}) and (\ref{Action1}) lead to time evolution equations for every field in equation (\ref{field1}) which reduce to equations (\ref{ELeqs}) when initial conditions $\Omega_1=\Omega_g=\Omega_s=0$ are chosen.  Showing this is tedious.

Thus, equations (\ref{Symplectic1}) and (\ref{Action1}) with variables (\ref{field1}) yield a Hamiltonian description of the dynamics if one can enforce $\Omega_1=\Omega_g=\Omega_s=0$, which requires writing the fields (\ref{field1}) in the action in terms of the constraints.  However, this solution is not ideal because the action does not have the standard symplectic form: the fields $A_i$ and $\pi^i$ appear in multiple terms in equation (\ref{Symplectic1}) via their presence in the constraints.  Unless the action is in standard symplectic form with the constraints themselves being fields in the argument list, it will be difficult to find the Poisson brackets and to eliminate the constraints.  Barcelos-Neto and Wotzasek provide a formal procedure for finding the Poisson brackets in this case.  We proceed differently, by changing variables to a good canonical set, following Maskawa and Nakajima. This is yet another way our work departs from the original Faddeev-Jackiw algorithm.

The method is based on using the constraints to replace other fields in the action.  We recall the first principle stated in Sec.~\ref{new}: we are free to transform the fields $\xi_a$ appearing in the action provided that the new fields span the same function space.  Thus, for each constraint added to the list of fields we must remove a field in such a way that the fields still span the same function space, and we must change the action accordingly.  This is straightforward, as we now show.

We begin with $\Omega_s$, which can be used to replace $A_0$ from the function list by solving the constraint (\ref{Omegas}), yielding
\begin{equation}\label{A0constr}
  A_0=\Delta^{-1}\left(\frac{\Omega_s}{\sqrt{\gamma}}\right)-{\cal\tilde D}^i
    \left[\frac{\alpha}{\sqrt{\gamma}}\gamma_{ij}\pi^j-\beta^k(\partial_iA_k-\partial_kA_i)
    +\frac{1}{\sqrt{\gamma}}\gamma_{ij}\partial_t\left(\sqrt{\gamma}\gamma^{jk}\right)
    A_k\right]
\end{equation}
where $\Delta^{-1}$ and ${\cal\tilde D}^i$ are defined in the Appendix.  Where $A_0$ appears in the action --- only in the Hamiltonian $H^0$ --- we replace it by the combination of $\Omega_s$, $A_i$ and $\pi^i$ given by equation (\ref{A0constr}).  This makes no change in the symplectic action.

Using the transverse-longitudinal decomposition (\ref{longtrans1}), equations (\ref{Omega1}) and (\ref{Omegag}) allow us to replace the longitudinal parts of $A_i$ and $\pi^i$ with the constraints $\Omega_g$ and $\Omega_1$,
\begin{equation}\label{Apiconstr}
  A_i={\cal P}_i^{\ \,j}\hat A_j+\partial_i\left[\Delta^{-1}\left(
    \frac{\Omega_g}{\sqrt{\gamma}}\right)\right]\ ,\ \ 
  \pi^i={\cal P}^i_{\ j}\hat\pi^j+{\cal D}^i(\Omega_1-\rho)\ .
\end{equation}

Projection operators (defined in the Appendix) eliminate the longitudinal parts of $\hat A_i$ and $\hat\pi^i$ so that they contribute only to the transverse parts of the fields.  The longitudinal parts of $A_i$ and $\pi^i$ are represented by ${\cal\tilde D}^iA_i=\Delta^{-1}(\Omega_g/\sqrt{\gamma})$ and $\partial_i\pi^i=\rho-\Omega_1$; for fixed values of the other fields they are completely determined by $\Omega_g$ and $\Omega_1$.  Therefore the fields $(\hat A_i,\hat\pi^i,\Omega_g,\Omega_1)$ span the same function space as $(A_i,\pi^i)$ and are a satisfactory replacement for them, given equations (\ref{Apiconstr}).

Equations (\ref{A0constr}) and (\ref{Apiconstr}) allow us to change our list of variables in the action to
\begin{equation}\label{field1a}
  \xi_a\in\left\{\hat A_i,\hat \pi^i,\phi_1,\phi_2,\pi_1,\pi_2,\lambda_1,\Omega_1,\lambda_g,\Omega_g,\lambda_s,\Omega_s\right\}\ .
\end{equation}
It can be shown that equations  (\ref{Symplectic1}) and (\ref{Action1}) with (\ref{Apiconstr}) lead to time evolution equations for every field in equation (\ref{field1a}) which reduce to equations (\ref{ELeqs}) when initial conditions $\Omega_1=\Omega_g=\Omega_s=0$ are chosen.  Showing this is much less onerous than it was for the set of fields (\ref{field1}) but will not be done here.

Although equations  (\ref{Symplectic1}) and (\ref{Action1}) produce time evolution equations for every field in (\ref{field1a}), the action is still not in the standard symplectic form because $(\phi_1,\phi_2,\pi_1,\pi_2,\Omega_1,\Omega_g)$ appear in multiple terms in equation (\ref{Symplectic1}). This makes it difficult to find Poisson brackets.  They will be found easily once we put the action in standard symplectic form by redefining the variables once more.

There are many possible choices of variables that put the action in standard symplectic form, all of them related by canonical transformations.  A particularly convenient choice follows from
\begin{equation}\label{redefine}
  \pi^i_\perp\equiv{\cal P}^i_{\ j}\hat\pi^j\ ,\ \ 
  \lambda_g'=\lambda_g+\Delta^{-1}\left(\frac{\Omega_1-\rho}{\sqrt{\gamma}}\right)
\end{equation}
so that
\begin{eqnarray}\label{pidotAi}
  \pi^i\dot A_i+\dot\lambda_g\Omega_g
  &=&\pi^i_\perp(\partial_t\hat A_i)+\dot\lambda_g'\Omega_g\nonumber\\
  &&-\hat A_i\dot{\cal D}^i(\rho-\Omega_1)+\left[\Delta^{-1}\left(\frac{\Omega_g}
    {\sqrt{\gamma}}\right)\right]\partial_t\left(\sqrt{\gamma}\Delta\right)
    \left[\Delta^{-1}\left(\frac{\rho-\Omega_1}{\sqrt{\gamma}}\right)\right]
\end{eqnarray}
plus boundary terms from integration by parts that have been dropped.  Two new terms have appeared that will be moved to the Hamiltonian because they involve no time derivatives of the fields in the action.

With the transformation (\ref{redefine}) we have a new set of fields
\begin{equation}\label{field2}
  \xi_a\in\left\{\hat A_i,\pi^i_\perp,\phi_1,\pi_1,\phi_2,\pi_2,\lambda_1,\Omega_1,\lambda_g',\Omega_g,\lambda_s,\Omega_s\right\}
\end{equation}
that places the action in perfect symplectic form, $S=I^2-\int dt\,H^2$ with
\begin{equation}\label{Symplectic2}
  I^2=\int d^4x\left(\pi^i_\perp\partial_t\hat A_i+\pi_1\dot\phi_1+\pi_2\dot\phi_2
    +\dot\lambda_1\Omega_1+\dot\lambda_g'\Omega_g+\dot\lambda_s\Omega_s\right)
\end{equation}
and
\begin{equation}\label{Hamiltonian2}
  H^2[\xi]=H^0+\int d^3x\left\{\hat A_i\dot{\cal D}^i(\Omega_1-\rho)-\left[\Delta^{-1}
    \left(\frac{\Omega_g}{\sqrt{\gamma}}\right)\right]\partial_t\left(\sqrt{\gamma}\Delta\right)
    \left[\Delta^{-1}\left(\frac{\Omega_1-\rho}{\sqrt{\gamma}}\right)\right]\right\}\ .
\end{equation}
In $H^0[A_0,A_i,\pi^i,\phi_1,\phi_2,\pi_1,\pi_2]$, the fields $A_0$, $A_i$ and $\pi^i$ must be expressed in terms of the fields (\ref{field2}) using equations (\ref{A0constr}) and (\ref{Apiconstr}), with $\pi^i_\perp={\cal P}^i_{\ j}\hat\pi^j$ so that
\begin{equation}\label{piperp}
  \partial_i\pi^i_\perp=0\ .\
\end{equation}

How can we restrict $\pi^i_\perp$ to be transverse?  It is natural to introduce a new constraint, $\Omega_2\equiv\partial_i\pi^i_\perp$,  with a new Lagrange multiplier $\dot\lambda_2$.  This will lead us back to the second of equations (\ref{Apiconstr}) with $\Omega_1$ replaced by $\Omega_1+\Omega_2$, without making visible progress.  Another way is to simply declare that the Hilbert space of fields $\pi^i_\perp$ is restricted to transverse fields satisfying (\ref{piperp}), as is implied by the first of equations (\ref{redefine}).  At the same time, $\hat A_i$ can also be restricted to transverse fields obeying $\nabla^i\hat A_i=0$.  If it is inconvenient to restrict the Hilbert space to transverse fields, then $\pi^i_\perp$ can be replaced by ${\cal P}^i_{\ j}\pi^j_\perp$ everywhere that it appears in the Hamiltonian (but not in the symplectic action).

From the symplectic action (\ref{Symplectic2}) in standard form it is easy to obtain Poisson brackets, as they are defined by
\begin{equation}\label{PBdef}
  \dot\xi_a({\bm x},t)\equiv\int d^3y\frac{\delta I}{\delta\xi_b({\bm y},t)}
    \Bigl\{\xi_a({\bm x},t),\xi_b({\bm y},t)\Bigr\}
\end{equation}
where $\xi_a$ and $\xi_b$ are any of the fields in the action. Notably, there are no additional source terms such as the one that appears in (\ref{Hamfield}). Poisson brackets are always evaluated at equal times, so from now on the time argument will be dropped.  Poisson brackets are
 anticommuting
\begin{equation}\label{anticomm}
  \Bigl\{A({\bm x}),B({\bm y})\Bigr\}=-\Bigl\{B({\bm y}),A({\bm x})\Bigr\}\ ,
\end{equation}
they obey the Leibniz rule
\begin{equation}\label{leibniz}
  \Bigl\{A({\bm x})B({\bm x}),C({\bm y})\Bigr\}=\Bigl\{A({\bm x}),C({\bm y})\Bigr\}
    B({\bm x})+A({\bm x})\Bigl\{B({\bm x}),C({\bm y})\Bigr\} \ ,
\end{equation}
and they are bilinear,
\begin{eqnarray}\label{bilinear}
  &&\Bigl\{a_1({\bm x})A_1(x)+a_2({\bm x})A_2({\bm x}),
    b_1({\bm y})B_1({\bm y})+b_2({\bm y})B_2({\bm y})\Bigr\}
    =a_1b_1\Bigl\{A_1({\bm x}),B_1({\bm y})\Bigr\}\nonumber\\
  &&+a_1b_2\Bigl\{A_1({\bm x}),B_2({\bm y})\Bigr\}
    +a_2b_1\Bigl\{A_2({\bm x}),B_1({\bm y})\Bigr\}
    +a_2b_2\Bigl\{A_2({\bm x}),B_2({\bm y})\Bigr\}\ ,
\end{eqnarray}
where $a_i({\bm x})$ and $b_i({\bm y})$ are any linear differential or integral operators.

The nonzero Poisson brackets can be read off directly from equation (\ref{Symplectic2}) using  (\ref{PBdef}):
\begin{eqnarray}\label{GLPB}
    \Bigl\{\hat A_i({\bm x}),\pi^j_\perp({\bm y})\Bigr\}
    &=&\delta_i^{\ j}\delta^3({\bm x}-{\bm y})\ ,\nonumber\\
  \Bigl\{\phi_1({\bm x}),\pi_1({\bm y})\Bigr\}
  &=&\delta^3({\bm x}-{\bm y})\ ,\nonumber\\
  \Bigl\{\phi_2({\bm x}),\pi_2({\bm y})\Bigr\}
  &=&\delta^3({\bm x}-{\bm y})\ ,\nonumber\\
  \Bigl\{\lambda_1({\bm x}),\Omega_1({\bm y})\Bigr\}
  &=&\delta^3({\bm x}-{\bm y})\ ,\nonumber\\
  \Bigl\{\lambda_g'({\bm x}),\Omega_g({\bm y})\Bigr\}
    &=&\delta^3({\bm x}-{\bm y})\ ,\nonumber\\
  \Bigl\{\lambda_s({\bm x}),\Omega_s({\bm y})\Bigr\}
  &=&\delta^3({\bm x}-{\bm y})\ .
\end{eqnarray}
The first Poisson bracket implies that $\hat A_i$ and $\pi^j_\perp$ are allowed to range over the complete set of one-form and vector fields, not merely transverse fields: there is a Kronecker delta function, not a transverse projection factor, in the equation, so that all field components are equally valid.  As noted previously, one way to handle this is to replace $\pi^i_\perp$ by ${\cal P}^i_{\ j}\pi^j_\perp$.  Equivalently, we can replace the standard Poisson bracket for $\hat A_i$ and $\pi^j_\perp$ by the transverse Poisson bracket defined by
\begin{equation}\label{PBtrans}
    \Bigl\{\hat A_i({\bm x}),\pi^j_\perp({\bm y})\Bigr\}_\perp
    ={\cal P}^j_{\ i}({\bm y})\delta^3({\bm x}-{\bm y})={\cal P}_i^{\ \,j}({\bm x})
    \delta^3({\bm x}-{\bm y})\ .
\end{equation}
Longitudinal fields make no contribution to the transverse bracket, making it a convenient way to work in the restricted function space for the transverse fields. The presence of the projection operator identifies the part of function space that the fields live on.

Incidentally, the Poisson brackets allow us to understand the difficulties of second-class constraints in the Dirac formalism and in the Faddeev-Jackiw iteration, where the constraint $\Omega_s$ would not have been found after $\Omega_1$ was added to the symplectic action.  The reason why can be seen using equations (\ref{redefine}) and (\ref{GLPB}), from which $\{\lambda_1,\lambda_g\}\ne0$.  Because $\lambda_g$ generates $\Omega_s=\dot\Omega_g$, the presence of both $\lambda_1$ and $\lambda_g$in the action eliminates the zero-mode of the symplectic action corresponding to $\Omega_s$, so it cannot be found using the Faddeev-Jackiw method if $\dot\lambda_1\Omega_1$ has first been added to the action.  To find $\Omega_s$  one would have to first remove $\dot\lambda_1\Omega_1$ from the symplectic action, a dubious procedure unless $\Omega_1$ is also removed from the Hamiltonian.  It is much easier to find $\Omega_s$ by direct differentiation of $\Omega_g$.

The advantage of the method presented above compared with the Dirac method is that one never has to start with the wrong brackets, nor must one introduce weak equality.  Classical equations of motion are used repeatedly to find constraints, which is appropriate because constraints reside in the part of Hilbert space that will be restricted to solutions of Hamilton's equations.  Compared with the Faddeev-Jackiw method, the procedure given in this section is simpler because one never has to calculate inverse brackets.

\subsection{Toward a Fully Reduced Hamiltonian}\label{GLA4}

The final step is to eliminate the constraints.  This follows from treating the constraints and their Lagrange multipliers as classical fields obeying Hamilton's equations of motion.  Hamilton's equations of motion are simply
\begin{equation}\label{Hameqs}
  \dot\xi_a({\bm x})=\Bigl\{\xi_a({\bm x}),H\Bigr\}
    =\int d^3y\frac{\delta H}{\delta\xi_b({\bm y})}\Bigl\{\xi_a({\bm x}),\xi_b({\bm y})\Bigr\}
\end{equation}
where $\xi_a$ is any of the fields in the symplectic action.  For a functional $F[\xi,t]$ of the fields and separately of time (through time-dependence of external fields such as the metric fields $\alpha$, $\beta_i$ and $\gamma_{ij}$),
\begin{equation}\label{Fdot}
  \dot F=\partial_tF+\{F,H\}=\partial_tF+\int d^3x\frac{\delta F}{\delta\xi_a({\bm x})}
    \Bigl\{\xi_a({\bm x}),H\Bigr\}\ .
\end{equation}

Applying this to the constraints and Lagrange multipliers gives
\begin{eqnarray}\label{constrevol}
  \dot\Omega_1&=&-\frac{\delta H}{\delta\lambda_1}=0\ ,\nonumber\\
  \dot\Omega_g&=&-\frac{\delta H}{\delta\lambda_g'}=0\ ,\nonumber\\
  \dot\Omega_s&=&-\frac{\delta H}{\delta\lambda_s}=0\ ,\nonumber\\    \sqrt{\gamma}\Delta\dot\lambda_1&=&\sqrt{\gamma}\Delta\left(\frac{\delta H}
    {\delta\Omega_1}\right)=\Omega_s+\partial_i\left(\alpha{\cal D}^i\Omega_1\right)
    \ ,\nonumber\\
  \sqrt{\gamma}\Delta\dot\lambda_g&=&-\partial_t\left(\sqrt{\gamma}\Delta\right)
    \left[\Delta^{-1}\left(\frac{\Omega_1}{\sqrt{\gamma}}\right)\right]\ ,\nonumber\\
  \sqrt{\gamma}\Delta\dot\lambda_s&=&\sqrt{\gamma}\Delta\left(\frac{\delta H}
    {\delta\Omega_s}\right)=\Omega_1\ .
\end{eqnarray}
Instead of the equation of motion for $\lambda_g'$ we give above the equation of motion for $\lambda_g$.  It is worth showing how $\dot\lambda_g$ is computed.  Using (\ref{redefine}),
\begin{equation}\label{dlg1}
  \partial_t\left(\sqrt{\gamma}\Delta\lambda_g\right)=
    \partial_t\left(\sqrt{\gamma}\Delta\lambda_g'\right)+\Bigl\{\rho-\Omega_1,H\Bigr\}\ .
  \nonumber\\
\end{equation}
Now using (\ref{Fdot}),
\begin{eqnarray}
  \sqrt{\gamma}\Delta\dot\lambda_g&=&
  \sqrt{\gamma}\Delta\dot\lambda_g'+\partial_t\left(\sqrt{\gamma}\Delta\right)(\lambda_g'
    -\lambda_g)+\Bigl\{\rho-\Omega_1,H\Bigr\}\nonumber\\
  &=&\sqrt{\gamma}\Delta\left(\frac{\delta H}{\delta\Omega_g}\right)
    +\partial_t\left(\sqrt{\gamma}\Delta\right)\left[\Delta^{-1}\left(\frac{\rho-\Omega_1}
    {\sqrt{\gamma}}\right)\right]+\Bigl\{\rho-\Omega_1,H\Bigr\}\nonumber\\
  &=&-\partial_i\left(\frac{\delta H^0}{\delta A_i}\right)+\Bigl\{\rho-\Omega_1,H\Bigr\}
    -\partial_t\left(\sqrt{\gamma}\Delta\right)\left[\Delta^{-1}\left(\frac{\Omega_1}
    {\sqrt{\gamma}}\right)\right]\ .
\end{eqnarray}
The functional derivative $\delta H^0/\delta A_i=-\dot\pi^i$ is given by the second of equations (\ref{ELeqs}).  Thus, using (\ref{Omega1}), we obtain
\begin{equation}\label{Omega1dot}
  -\partial_i\left(\frac{\delta H^0}{\delta A_i}\right)+\Bigl\{\rho-\Omega_1,H\Bigr\}
  =\partial_t\partial_i\pi^i+\Bigl\{\rho-\Omega_1,H\Bigr\}
  =\Bigl\{\partial_i\pi^i+\rho-\Omega_1,H\Bigr\}=0\ .
\end{equation}
The same result can be shown by substituting $\delta H^0/\delta A_i$ and using the Leibnitz rule for $\{\rho,H\}$ with $\rho=q(\phi_2\pi_1-\phi_1\pi_2)$.

Up to this point we have made no assumptions at all about the values of the constraints; in particular, we have not assumed that any of the constraint fields vanish.  These fields are treated like any others in the action.  Now consistency with the original action plus gauge-fixing requires $\Omega_1=\Omega_g=\Omega_s=0$.  Remarkably, if these conditions hold everywhere at any initial time, then equations (\ref{constrevol}) show that they hold at all times, provided that they and the Lagrange multipliers are governed by Hamilton's equations.  Moreover, we see that the Lagrange multipliers are all constants, which may be set to zero.  One notable exception is $\lambda_g'$ which depends on $\rho$ in equation (\ref{redefine}).  Since this field does not appear in the Hamiltonian and it has vanishing Poisson bracket with all fields except $\Omega_g$ which itself vanishes, we may ignore $\lambda_g'$ in the dynamics.

It is worth commenting on the assumption that the constraints and Lagrange multipliers are treated as classical fields obeying Hamilton's equations.  In quantum field theory, the same is not true of the remaining fields in the action, e.g. $\hat A_i$.  Those fields fluctuate away from the classical solution.  Constraints are not physical fields; their presence indicates that there are fewer physical degrees of freedom than fields in the original action.  Only by proceeding as we have done can the physical and constrained fields be separated.

As a result, the constraints are dynamically enforced, and they can be set to zero inside the action.  The result is a final reduction of the action to
\begin{equation}\label{Symplectic3}
  S[\hat A_i,\pi^i_\perp,\phi_1,\pi_1,\phi_2,\pi_2]=\int d^4x\left(\pi^i_\perp\partial_t\hat A_i
    +\pi_1\dot\phi_1+\pi_2\dot\phi_2\right)-H\ ,\ \ 
\end{equation}
with
\begin{eqnarray}\label{Hamiltonian3}
  H&=&\int d^3x\Biggl\{\frac{\alpha}{2\sqrt{\gamma}}\gamma_{ij}\pi^i\pi^j
    +\beta^i\pi^jF_{ij}+\frac{1}{4}\alpha\sqrt{\gamma}\gamma^{ik}\gamma^{jl}F_{ij}F_{kl}
    +\frac{\alpha}{2\sqrt{\gamma}}\left(\pi_1^2+\pi_2^2\right)\nonumber\\
  &&+\beta^i\left(\pi_1\partial_i\phi_1+\pi_2\partial_i\phi_2\right)
    +A_i(\dot{\cal D}^i-\beta^i)\rho
    +\frac{\alpha\sqrt{\gamma}}{2}\left[-\mu^2\left(\phi_1^2+\phi_2^2\right)
    +\frac{1}{2}\lambda\left(\phi_1^2+\phi_2^2\right)^2\right]\nonumber\\
  &&+\frac{\alpha\sqrt{\gamma}}{2}\gamma^{ij}\left[(\partial_i\phi_1+qA_i\phi_2)    
    (\partial_j\phi_1+qA_j\phi_2)+(\partial_i\phi_2-qA_i\phi_1)(\partial_j\phi_2-qA_j\phi_1)
    \right]\Biggr\}
\end{eqnarray}
and
\begin{equation}\label{finalfields}
  A_i\equiv{\cal P}_i^{\ \,j}\hat A_j\ ,\ \  \pi^i=\pi^i_\perp+{\cal D}^i\rho\ ,\ \ 
  \rho\equiv q(\phi_1\pi_2-\phi_2\pi_1)\ .
\end{equation}
Note that $A_0$ has dropped out of the Hamiltonian after the enforcement of $\Omega_1=0$; it can be found from the other fields using equation (\ref{A0constr}) with $\Omega_s=0$.  In the Hamiltonian (\ref{Hamiltonian3}), $\pi^i_\perp$ should be restricted to the space of transverse fields. This defines the previously mentioned function space.

The final check is to see whether equations (\ref{Symplectic3})--(\ref{finalfields}) give the correct classical equations of motion for $A_i$ and $\pi^i$.  The check is simply to differentiate the fields using equation (\ref{Fdot}):
\begin{eqnarray}\label{Aidot}
  \dot A_i&=&{\cal P}_i^{\ \,j}\partial_t\hat A_j+\dot{\cal P}_i^{\ \,j}\hat A_j\nonumber\\
  &=&{\cal P}_i^{\ \,j}\Bigl\{\hat A_j,H\Bigr\}-\partial_i\left(\dot{\cal\tilde D}^j\hat A_j
    \right)\nonumber\\
  &=&{\cal P}_i^{\ \,j}\frac{\delta H}{\delta\pi^j_\perp}-\partial_i\left(\dot{\cal\tilde D}^j
    A_j\right)\nonumber\\
  &=&{\cal P}_i^{\ \,j}\frac{\delta H^0}{\delta\pi^j}-\partial_i\partial_t\left({\cal\tilde D}^j
    A_j\right)+\partial_i\left({\cal\tilde D}^j\dot A_j\right)\ .
\end{eqnarray}
The gauge condition $\Omega_g=0$ enforces ${\cal\tilde D}^jA_j=0$ so equation (\ref{Aidot}) reduces to
\begin{equation}\label{AidotP}
  {\cal P}_i^{\ \,j}\dot A_j={\cal P}_i^{\ \,j}\frac{\delta H^0}{\delta\pi^j}
\end{equation}
which is the transverse part of the original equation of motion; the longitudinal part is enforced by $\Omega_g=0$.  Similarly,
\begin{eqnarray}\label{pidot}
  \dot\pi^i&=&\dot\pi^i_\perp+{\cal D}^i\dot\rho+\dot{\cal D}^i\rho\nonumber\\
  &=&\Bigl\{\pi^i_\perp,H\Bigr\}+{\cal D}^i\Bigl\{\rho,H\Bigr\}+\dot{\cal D}^i\rho
    \nonumber\\
  &=&-\frac{\delta H}{\delta\hat A_i}+{\cal D}^i\Bigl\{\rho,H\Bigr\}+\dot{\cal D}^i\rho
    \nonumber\\
  &=&-{\cal P}^i_{\ j}\frac{\delta H^0}{\delta A_j}+{\cal D}^i\Bigl\{\rho,H\Bigr\}\ .
\end{eqnarray}
The constraint $\Omega_1=0$ gives
\begin{equation}\label{pilong}
  \partial_i\pi^i=\rho\ \Rightarrow\ \partial_i\dot\pi^i=\dot\rho=\Bigl\{\rho,H\Bigr\}
\end{equation}
yielding
\begin{equation}\label{pidotP}
  {\cal P}^i_{\ j}\dot\pi^j=\dot\pi^i-{\cal D}^i(\partial_j\dot\pi^j)=-{\cal P}^i_{\ j}
    \frac{\delta H^0}{\delta A_j}
\end{equation}
which is the transverse part of the original equation of motion.  The longitudinal part is enforced by the constraint $\Omega_1=0$.  Notice that the correct equations of motion follow whether we use the original Poisson bracket $\{\hat A_i({\bm x}),\pi^j({\bm y})\}=\delta_i^{\ \,j}
\delta^3({\bm x}-{\bm y})$ or the transverse Poisson bracket (\ref{PBtrans}).

We have obtained the complete Hamiltonian reduction of the Ginzburg-Landau model in curved spacetime.  The canonical fields $\hat A_i$ and $\pi^i_\perp$ are related to the physical fields by equations (\ref{finalfields}).  The bare Hamiltonian gets supplemented by a curved spacetime contribution in equation  (\ref{Hamiltonian3}).  We have carefully analyzed the Hilbert space of fields.  As an application, it will be interesting to look at the quantum field theory of this model in black hole spacetimes, to see how Hawking radiation arises in canonical quantization.

\section{Conclusion}\label{conclusion}
We have presented a reformulation of the Faddeev-Jackiw algorithm for computing a Hamiltonian and Poisson brackets from a singular Lagrangian. In particular, inspired by the Barcelos-Neto-Wotzasek reformulation of the algorithm for discrete actions, we implement a nearly analogous approach for the continuous case, fields. We show it is possible to eliminate the need for calculating the individual components of the symplectic form, as is done in the BNW procedure. We emphasize the importance of properly partitioning the phase space as a means to finding the correct symplectic picture. To show how this new approach works, we apply use it to find to the Hamiltonian and Poisson brackets for the Ginzburg-Landau action in a curved spacetime, which has not previously appeared in the literature.

This new technique is of interest because of its prospective application to gauge theories, whose Lagrangians are singular. It eliminates the need to use the sometimes confusing Dirac procedure to calculate the Poisson brackets associaed with a Lagrangian. It therefore provides an alternative to the BRST~\citep{1976AnPhy..98..287B} approach to quantizing gauge theories. 

Prospective applications include a study of the behavior of axionic dark matter in a Friedmann-Robertson-Walker background. This question is of particular interest because of Sikivie and Yang's~\citep{2009PhRvL.103k1301S} proposal that axionic dark matter gravitationally thermalizes to form a Bose-Einstein condensate. Additional formal steps include an application to fermionic fields as well as the coupling of matter fields to General Relativity where the metric is treated dynamically.

\subsection*{Acknowledgements}
C.P. would like to acknowledge helpful comments and discussions with Roman Jackiw, Alan Guth, David Kaiser, Janet Conrad, Peter Fisher, Wesley Harris, Margaret Edwards, Tehani Finch, Jolyon Bloomfield, Mark Hertzberg and Meng Su. C.P. is supported by the M.I.T. Martin Luther King, Jr. Visiting Professors and Scholars program.
\appendix
\section*{Appendix: Differential operators in curved spaces}
\label{sec:Appendix}

In the 3+1 split of curved spacetime used in these notes we have frequent need for spatial derivative operators obtained from the metric $\gamma_{ij}$ in addition to $\nabla_i$, the spatial covariant derivative with respect to the metric $\gamma_{ij}$.  The needed tools are summarized in this appendix.

The 3-divergence of a one-form and the 3-Laplacian of a scalar are
\begin{equation}\label{diver3}
  \nabla_iA^i=\nabla^iA_i=\frac{1}{\sqrt{\gamma}}\partial_i\left(\sqrt{\gamma}
    \gamma^{ij}A_j\right)\ ,\ \ 
  \Delta f\equiv\nabla_i\nabla^if=\frac{1}{\sqrt{\gamma}}\partial_i\left(\sqrt{\gamma}
    \gamma^{ij}\partial_jf\right)\ .
\end{equation}
We will also use the inverse Laplacian operator $\Delta^{-1}$, which is defined by the solution of the Poisson equation for scalar fields $f$ and $g$,
\begin{equation}\label{Poisson}
  \Delta f=g\ \ \to\ \ f=\Delta^{-1}g\ ,
\end{equation}
assuming the existence of suitable boundary conditions (e.g. $f\to0$ at infinity for a localized source in an infinite space).

Integration at fixed time over a spatial volume $V$ with boundary ${\partial V}$, the
Laplacian obeys the following useful identity for scalar fields $f$ and $g$,
\begin{equation}\label{Laplaceint}
  \int_V d^3x\,f\left(\sqrt{\gamma}\Delta\right)g
  =\int_V d^3x\,g\left(\sqrt{\gamma}\Delta\right)f
    +\int_{\partial V}dS_i\,\sqrt{\gamma}\gamma^{ij}\left(f\partial_jg-g\partial_jf\right)\ ,
\end{equation}
where $dS_i$ is the suitably oriented coordinate area element, e.g. $dS_r=d\theta d\phi$ for a spherical boundary in flat spacetime.  The factors of $\sqrt{\gamma}$ correct coordinate volumes to proper volumes.  These factors can be absorbed into scalar fields, converting them to scalar densities, which we generically denote by $\rho$ or $\pi$.

When using curvilinear coordinates it is necessary to distinguish between spatial scalar fields and scalar densities because the canonical momentum of a scalar field is a scalar density.  Thus, $\phi_1$ and $\phi_2$ are scalar fields while $\pi_1$, $\pi_2$, and the Lagrangian density ${\cal L}$ are scalar densities.  This is evident from the absence of factors $\sqrt{\gamma}$ preceding $\pi\dot\phi$ and ${\cal L}$ in the action.  The constraints $\Omega_1$, $\Omega_g$ and $\Omega_s$ are all scalar densities.  Similarly, the momentum conjugate to the one-form field $A_i$, $\pi^i$, is a vector density; it is converted to a spatial vector by dividing by a factor $\sqrt{\gamma}$.

Integration by parts is used repeatedly in the Hamiltonian formulation giving rise to boundary terms as in equation (\ref{Laplaceint}).  Such terms in the action make no contribution to functional derivatives within a boundary although they do affect boundary conditions on the field equations.  We neglect these terms throughout.  If the spatial region has a boundary, then a more careful analysis is required to evaluate boundary effects.

It is convenient to introduce spatial inverse differential operators that act on spatial scalar densities and one-forms respectively,
\begin{equation}\label{invDiff}
  {\cal D}^i\rho\equiv\sqrt{\gamma}\gamma^{ij}\partial_j\left[\Delta^{-1}\left(\frac{\rho}
    {\sqrt{\gamma}}\right)\right]\ ,\ \ 
  \tilde{\cal D}^ih_i\equiv\Delta^{-1}\left[\frac{1}{\sqrt{\gamma}}\partial_j
    \left(\sqrt{\gamma}\gamma^{ij}h_i\right)\right]\ ,
\end{equation}
which have the properties
\begin{equation}\label{DiffInv}
  \partial_i\left({\cal D}^i\rho\right)=\rho\ ,\ \ \tilde{\cal D}^i\left(\partial_if\right)=f
\end{equation}
and
\begin{equation}\label{DiffIntswap}
  \int d^3x\,h_i{\cal D}^i\rho=-\int d^3x\,\rho\tilde{\cal D}^ih_i\ ,
\end{equation}
where possible boundary terms from integration by parts have been dropped in the last result.  Note that ${\cal D}^i\rho$ is a vector density so that no factors of $\sqrt{\gamma}$ are needed in equation (\ref{DiffIntswap}).

The inverse differential operators allow us to define transverse projection operators, which act on spatial one-forms and vector densities, respectively:
\begin{eqnarray}\label{projtrans}
  {\cal P}_i^{\ \,j}A_j&\equiv&\left(\delta_i^{\ j}-\partial_i{\cal\tilde D}^j\right)A_j
  =A_i-\partial_i\Delta^{-1}\left[\frac{1}{\sqrt{\gamma}}
    \partial_k(\sqrt{\gamma}\gamma^{kj}A_j)\right]\ ,\\
  {\cal P}^i_{\ j}\pi^j&\equiv&\left(\delta^i_{\ j}-{\cal D}^i\partial_j\right)\pi^j
  =\pi^i-\sqrt{\gamma}\gamma^{ik}\partial_k\left[
    \Delta^{-1}\left(\frac{\partial_j\pi^j}{\sqrt{\gamma}}\right)\right]\ .\nonumber
\end{eqnarray}
These differential operators obey the following useful identities,
\begin{equation}\label{dproj}
  \partial_j{\cal P}^j_{\ i}={\cal P}_i^{\ \,j}\partial_j=0\ ,\ \ 
  {\cal P}^i_{\ j}{\cal D}^j={\cal\tilde D}^j{\cal P}_j^{\ \,i}=0,\ \ 
  {\cal P}^i_{\ j}{\cal P}^j_{\ k}={\cal P}^i_{\ k}\ ,\ \ 
  {\cal P}_i^{\ \,j}{\cal P}_j^{\ \,k}={\cal P}_i^{\ \,k}\ .
\end{equation}

As a consequence of equations (\ref{projtrans}) and (\ref{dproj}), arbitrary spatial one-forms and vector density fields can be decomposed into transverse and longitudinal parts as follows:
\begin{equation}\label{longtrans1}
  A_i={\cal P}_i^{\ \,j}\hat A_j+\partial_iA_\parallel\ ,\ \ 
  \pi^i={\cal P}^i_{\ j}\hat\pi^j+{\cal D}^i\pi_\parallel\ ,
\end{equation}
where
\begin{equation}\label{longtrans2}
  \hat A_i\equiv{\cal P}_i^{\ \,j}A_j\ ,\ \ A_\parallel\equiv{\cal\tilde D}^iA_i\ ,\ \ 
  \hat\pi^i\equiv{\cal P}^i_{\ j}\pi^j\ ,\ \ \pi_\parallel\equiv\partial_i\pi^i\ .
\end{equation}

Finally, we have the following time derivatives of the inverse differential operators:
\begin{eqnarray}\label{dotDelta}
  \dot{\cal D}^i\rho&\equiv&\partial_t({\cal D}^i\rho)-{\cal D}^i\dot \rho={\cal P}^i_{\ j}
    \left[\frac{\gamma_{kl}}{\sqrt{\gamma}}\partial_t\left(\sqrt{\gamma}\gamma^{jk}
    \right){\cal D}^l\rho\right]\ ,\ \ \nonumber\\
  \dot{\cal\tilde D}^ih_i&\equiv&\partial_t({\cal\tilde D}^ih_i)-{\cal\tilde D}^i\dot h_i
    ={\cal\tilde D}^i\left[\frac{\gamma_{ij}}{\sqrt{\gamma}}\partial_t\left(\sqrt{\gamma}
    \gamma^{jk}\right){\cal P}_k^{\ \,l}h_l\right]\ ,\ \ \nonumber\\
  \partial_t\left(\sqrt{\gamma}\Delta\right)f&\equiv&\partial_i\left[\partial_t\left(\sqrt{\gamma}
    \gamma^{ij}\right)\partial_jf\right]\ .
\end{eqnarray}
Combining equations (\ref{dproj}) and (\ref{dotDelta}) gives
\begin{equation}\label{ddot}
  \partial_i\dot{\cal D}^i\rho=0\ ,\ \ \dot{\cal\tilde D}^i\partial_if=0\ .
\end{equation}

\bibliography{FJtechnique_4}

\end{document}